\documentclass[useAMS,usenatbib,usegraphicx]{mn2e}
\usepackage{times}

\newcommand{\simlt}{\lower.5ex\hbox{$\; \buildrel < \over \sim \;$}}
\newcommand{\simgt}{\lower.5ex\hbox{$\; \buildrel > \over \sim \;$}}

\newcommand{\be}{\begin{equation}}
\newcommand{\ba}{\begin{eqnarray}}
\newcommand{\ee}{\end{equation}}
\newcommand{\ea}{\end{eqnarray}}

\title[Using bright ellipticals as dark energy cosmic clocks]
      {Using bright ellipticals as dark energy cosmic clocks}
\author[I.~Ferreras, A.~Melchiorri and D.~Tocchini-Valentini]
{Ignacio Ferreras$^1$, Alessandro Melchiorri$^{1,2}$ and 
Domenico Tocchini-Valentini$^1$\thanks{ferreras,melch,dtv@astro.ox.ac.uk}\\ 
1. Physics Dept. Denys Wilkinson Building, University of Oxford, 
Keble Road, Oxford OX1 3RH\\
2. Dipartimento di Fisica "G. Marconi", Universit\'a di Roma "La Sapienza",
   Ple Aldo Moro 2, 00185 Rome, Italy}
\begin{document}

\date{Draft version \today}

\pagerange{\pageref{firstpage}--\pageref{lastpage}} \pubyear{2003}

\maketitle

\label{firstpage}

\begin{abstract}
The age of the Universe has been increasingly constrained
by different techniques such as the observations
of type~Ia supernovae at high redshift or dating the
stellar populations of globular clusters. In this paper we 
present a complementary approach using the colours of the
brightest elliptical galaxies in clusters over a wide redshift
range ($z\simlt 1$). We put new and independent bounds on the
dark energy equation of state parameterised by a constant 
pressure-to-density ratio $w_Q$ as well as by a parameter
($\xi$) which determines the scaling between the matter and 
dark energy densities. We find that accurate estimates of the
metallicities of the stellar populations in moderate and high
redshift cluster galaxies can pose stringent constraints on
the parameters which describe the dark energy equation
of state. Our results are in 
good agreement with the analysis of dark energy models 
using SNIa data as constraint. Accurate estimates
of the metallicities of stellar populations in cluster galaxies 
at $z\simlt 2$ will make this approach a powerful complement 
to studies of cosmological parameters using high redshift SNIa.
\end{abstract}



\section{Introduction}
The discovery that the evolution of the Universe may be 
dominated by an effective cosmological constant is one of 
the most remarkable findings of recent years. 
An exceptional opportunity is now opening up to decipher the nature 
of dark energy, to test the veracity of theories 
and to reconstruct the dark energy's profile using a wide variety 
of observations over a broad redshift range.

One energy candidate that could possibly explain the observations is 
a dynamical scalar ``quintessence'' field. In many cases these models 
alleviate the fine-tuning and coincidence problems carried by a
cosmological constant (Zlatev et al. 1999).
The most relevant observational evidence for quintessence 
is that its equation of state parameter, 
$w_{Q}=p_Q/\rho_Q$ can assume values different from $-1$, 
which is the expected value in the case of a cosmological constant. 
Moreover, since very little is known currently about the nature 
of dark energy, an interaction (aside from gravity) of quintessence 
with the remaining forms of matter could be expected 
as well (e.g. Wetterich 1995).

Other deviations from a cosmological constant can be
considered as, for example, a sound speed different from 
light or perturbations in the dark energy fluid.
However these aspects of quintessence affect mainly the growth
of perturbations and their detectability is therefore subjected 
to the assumption of a model of structure formation. In this work
we will be most interested in the effect that dark energy
causes on the background evolution. In particular, we show that an 
accurate determination of the age of the Universe can be used to 
obtain important and independent information on the dark energy 
component.
 
Discrepancies between age determinations
have long plagued cosmology. 
The situation has changed drastically in recent years:
Type~Ia supernova measurements (Perlmutter et al. 1999; Riess et al. 1998), 
the acoustic peaks in the CMB anisotropies (Netterfield et al. 2001), 
cosmochronology  of long-lived radioactive nuclei in old Population~II stars 
(Cayrel et al. 2001), just to name a few, are all consistent with
an age $t_0=14\pm1$~Gyr. 
Furthermore, the recent age determination of the oldest white 
dwarf in the globular cluster M4 by the Hubble Space Telescope, 
with $t_{wd}=12.0\pm0.7$ at $95 \%$ c.l. (Hansen et al. 2002), 
can be used to put the strongest lower limit to date on the 
age of the Universe.

It is therefore timely to investigate if the new age determinations
can be used to produce new constraints on dark energy.
These results could also prove very useful in breaking some of the 
degeneracies that arise in cosmological parameter estimations from other 
observational sources (e.g. Bean \& Melchiorri 2002; Maor et al. 2002).

Recently, Krauss \& Chaboyer (2001), have provided constraints on 
$w_Q$ by using new globular cluster age estimates. 
Our analysis will differ in two new fundamental aspects: first we will
use age estimates based on old stellar populations at moderate redshift. 
As in (Ferreras, Melchiorri \& Silk 2001) we will provide estimates 
on the age of the Universe via a comparison of population synthesis models
with the photometry of the oldest stellar populations in bright cluster 
ellipticals. Second, following the parametrization of 
Dalal et al. (2001) we will also investigate the possibility of an 
interaction besides gravity between dark energy and dark 
matter. 


\section{Models of Dark Energy and the age of the Universe.}
In the literature one can find many proposals of scalar field 
dark energy models, differing by the choices of the self-interacting 
potentials (e.g. Sahni 2002), resulting in equations of state that in 
general depend on time.
On the other hand, the age of the Universe is most influenced by the 
relatively late evolution history. This enables us to simplify the problem 
by assuming a constant equation of state for the dark energy. 
A parameter determination through a likelihood analysis would very 
probably have a few chances of detecting a relatively smooth variation of 
the equation of state.
This line of reasoning is very similar to the one usually expressed
in many studies of constraint from SN-Ia (Maor et al. 2002).  

In fact, as our results will
show, the SNe parameter degeneracy is significantly similar to our case.
Thus our work in a way offers an independent check of those findings. 
This has its own relevance due to 
the intense debate on whether all the important systematic errors have been
taken into account in SNe data reports.
In what follows the sub-indices $Q$, $C$ and $B$ indicate respectively the
contributions from dark energy-quintessence, cold dark matter (CDM) and 
baryons. The collective CDM and baryonic contribute will be represented by the 
label $M$. We will adopt a quite general phenomenological model in which 
the ratio of the dark energy and dark matter densities scale as a power law:
\be
\rho_Q\propto \rho_C a^\xi,
\ee
with $\xi$ kept constant (as in Dalal et al. 2001). It should be noted that 
this assumption generally involves an interaction besides gravity. 
The dark energy equation of state, $w_Q=p_Q/\rho_Q$, is held fixed 
as stated above. Three interesting cases deserve particular attention. 
Putting $\xi=3$ and $w_Q=-1$ gives the $\Lambda$CDM model. 
The condition $\xi=-3w_Q$ corresponds to decoupled quintessence.
Furthermore the choice $\xi=0$ implies 
models that may explain the coincidence problem through an identical 
scaling of the two energy densities (e.g. Tocchini-Valentini \& Amendola 2002, 
Amendola \& Tocchini-Valentini 2001). The baryons are assumed coupled only 
through gravity to the rest of the world.

\begin{figure}
\includegraphics[width=3.3in]{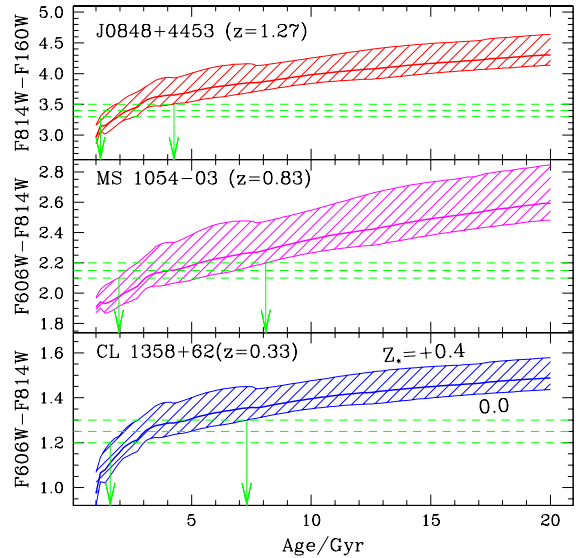}
\caption{Evolution with age of the colour of bright cluster galaxies 
at moderate redshift. The shading corresponds to metallicities in the 
range $0.0<Z_\star <0.4$~dex. The dashed lines represent the observed 
colour range of the brightest galaxies in the three clusters presented.
The arrows mark the allowed range at the observed redshift given the 
colour constraint (see text for details).}
\label{fig:colours}
\end{figure}

To perform a comparison with the data we will need an expression of the 
age of the Universe for this class of models. This is given by:
\be
t_0=9.8 {\rm Gyr} \ h^{-1}\int_0^1{\frac{da}{a} \frac{1}{\sqrt{f(a)}}}.
\label{eq:age}
\ee
If $\xi\neq0$ then
\be
\begin{array}{rcl}
f(a) &=& (\Omega_C + \Omega_Q)a^{-3} {\left[ 1- 
    \frac{\Omega_Q}{\Omega_C + \Omega_Q}(1-a^{\xi}) 
\right] } ^{-3w_Q/\xi}+\\
 & & +\Omega_B a^{-3}+(1-\Omega_C-\Omega_Q-\Omega_B)a^{-2}.
\end{array}
\ee
While when $\xi=0$ we have that
\be
f(a)= (\Omega_C + \Omega_Q)a^{-\beta}+ \Omega_B a^{-3}+
(1-\Omega_C-\Omega_Q-\Omega_B)a^{-2}
\ee
where
\be
\beta=3 \left( 1+ w_Q \frac{\Omega_Q}{\Omega_C+\Omega_Q} \right).
\ee
The inclusion of baryons is important since their contribution
may become significant not too far
in the past history of the Universe and their exclusion may overestimate 
the age of the Universe by more than $10 \%$ relative to the complete case, 
especially for highly negative values of $w_Q$.

In the decoupled case it has been shown that
values of the equation of state such that $w_Q<-1$ are not completely
excluded by observations (Melchiorri et al 2002, Schuecker et al. 2002).
 Therefore in our analysis we will consider the 
interval $-3<w_Q<0$. The limitations $0<\xi<9$ will be imposed on the
parameter $\xi$. To reduce the number of free parameters we consider a 
flat Universe. For models of coupled quintessence we fix $\Omega_M=0.3$ 
for the density of CDM and baryons altogether, and we also fix 
$\Omega_B=0.04$ for baryons.

\begin{figure}
\includegraphics[width=3.3in]{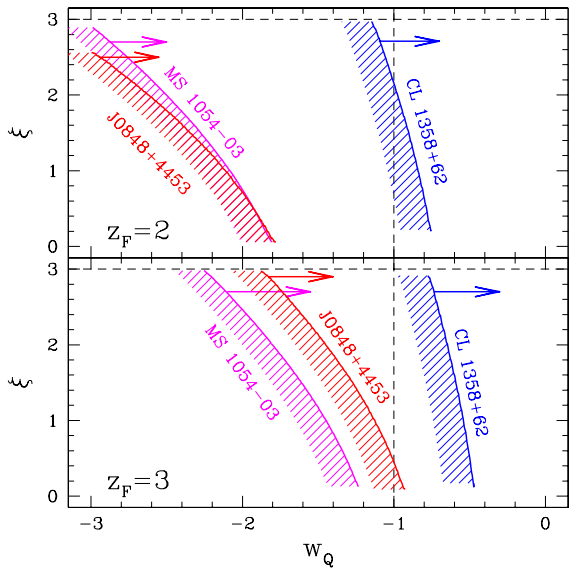}
\caption{Parameter space allowed by the colour constraints from the 
brightest ellipticals in the three clusters shown in 
figure~\ref{fig:colours}. Various cosmological parameters are fixed,
namely: $\Omega_M=0.3$; $H_0=72$km s$^{-1}$Mpc$^{-1}$ and $\Omega_B=0.04$.
The allowed region of parameter space stays to the right of the 
shaded line. We have chosen two of the latest formation redshifts 
compatible with the observations: $z_F=2$ ({\sl top panel}) and
$z_F=3$. Higher values of $z_F$ will further constrain
the allowed area of parameter space to the right.}
\label{fig:ages}
\end{figure}

\section{Using bright ellipticals as dark energy cosmic clocks}
The use of bright cluster ellipticals over a wide range of redshifts
allow us to set contraints on the age of the Universe. In the volume
spanned by the various cosmological parameters involved, these constraints 
go roughly ``in the same direction'' as the constraints from high redshift
SNe data --- which depend on the luminosity distance, and
thus provide an independent method to explore the allowed values of the
parameters. Ferreras et al. (2001) used this approach to
determine the age of the Universe and showed that the results were
consistent with estimates using other observations such as 
high redshift SNIa (Riess et al. 1998; Perlmutter et al. 1999), 
age dating of globular clusters (Salaris \& Weiss 1998; Pont et al. 1998;
Carretta et al. 2000) or nucleocosmochronology of old halo stars 
(Cayrel et al. 2001). 

Figure~\ref{fig:colours} shows the methodology of this approach. The colours
of the first few brightest ellipticals from clusters at moderate redshift
is compared with the predictions of the population synthesis models of 
Bruzual \& Charlot (1993) for a range of ages and metallicities. 
The latest available version of these models have been used. The shaded area
represents the colours for the metallicity range $0<Z_\star <+0.4$~dex, 
as expected in the central regions of bright ellipticals 
(Henry \& Worthey 1999). The dashed lines in each panel correspond
to the observed colours of the brightest ellipticals in clusters:
CL~1358+62 ($z=0.33$; Van Dokkum et al. 1998); MS~1054+03 ($z=0.83$; 
Van~Dokkum et al. 2000), and J0848+4453 ($z=1.27$; Van~Dokkum \& Stanford 2002).
The arrows give the allowed age of the Universe {\sl at the redshift
of the cluster} for the metallicity range expected in bright early-type
galaxies. The upper limit can be used to constrain the allowed values
of the parameters $w_Q$ and $\xi$. Figure~\ref{fig:ages} shows the 
constraint imposed by the upper limit expected for the three clusters 
shown in the previous figure. Some of the cosmological parameters
have been fixed, namely: $\Omega_M=0.3$, $\Omega_Q=1-\Omega_M$, 
$H_0=72$~km~s$^{-1}$~Mpc$^{-1}$, and $\Omega_B=0.04$. 
The allowed region stays to the right of the shaded
line for each of the three clusters. As reference, a $\Lambda$CDM
cosmology corresponds to $(w_Q,\xi )=(-1,3)$. A formation redshift of
$z_F\simgt 2$ is assumed, an arguably late epoch for the formation
of massive elliptical galaxies given the redshift evolution of the 
colour-magnitude relation (e.g. Stanford, Eisenhardt \& Dickinson 1998).
Earlier formation epochs are possible and will further constrain
the region of parameter space to the right of the figure.

A more comprehensive analysis was performed using the complete sample 
of 17 clusters from Stanford et al. (1998) over the redshift range
$0.3<z<0.9$. This sample was extracted from a larger sample of $46$
clusters drawn from a variety of optical, X-ray and radio-selected
clusters. The morphological classification was done from high resolution
images taken by the Wide Field and Planetary Camera~2 on board the
{\sl Hubble Space Telescope}. We selected the three brightest early-type
galaxies for each cluster, and used the published photometry in NIR
passbands $J$, $H$ and $K$, as well as on two optical filters which
straddle the 4000\AA\ break. Along similar lines as the method shown above,
we compare the photometry with the predictions of the latest 
population synthesis models of Bruzual \& Charlot (1993). Furthermore,
the stellar populations in these bright galaxies are assumed to 
evolve into those observed in bright ellipticals in local clusters such
as Coma or Virgo. Hence, a further constraint is imposed comparing the
$U-V$ colour when the galaxy is evolved to zero redshift. 

The models are run assuming a flat cosmology.
The dark energy parameters are explored over a large range: $-3<w_Q<0$;
and either $0<\xi<9$ or $0.1<\Omega_M<1$ for coupled and uncoupled 
models, respectively. The former assume fixed values for
$\Omega_M=0.3$ and $\Omega_B=0.04$. Uncoupled models are
obviously independent of the baryon density.
We assume a wide range of formation redshifts: $2<z_F<10$ and
stellar metallicities: $-0.1<Z_\star +0.3$. A Salpeter (1955) 
initial mass function (IMF) was used, although the results -- which 
are purely constrained by galaxy colour, not absolute luminosity --
are rather insensitive to the choice of IMF. 

\begin{figure}
\includegraphics[height=5.5in,width=3.4in]{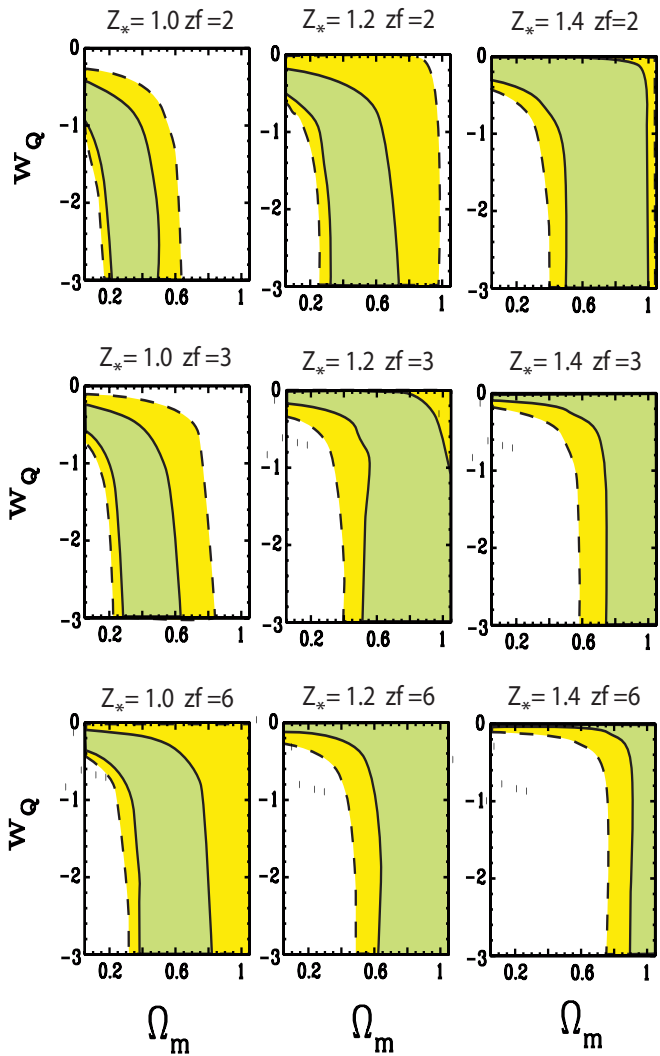}
\caption{Likelihood contours at the $68 \%$ (solid) and $95 \%$ (dashed) 
confidence levels in the parameter space
spanned by $\Omega_M$ and $w_Q$. 
Uncoupled dark energy models are considered. $Z_\star$ and $zf$
represent the metallicity (with respect to solar) and the 
formation redshift of the stellar populations, respectively.
}
\label{fig:quint}
\end{figure}

\section{Discussion}

\begin{figure}
\includegraphics[height=5.5in,width=3.4in]{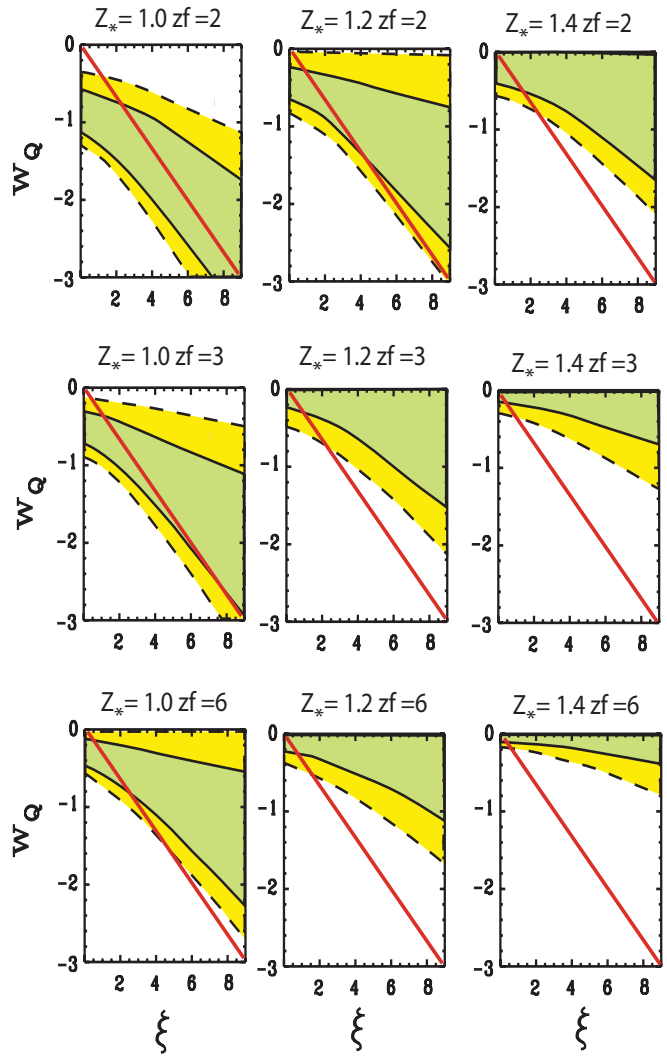}
\caption{Likelihood contours at the $68\%$ (solid) and 
$95\%$ (dashed) confidence levels in the parameter space
spanned by $\xi$ and $w_Q$ for coupled dark energy models.
All models in this figure assume a fixed $\Omega_M=0.3$.
$Z_\star$ and $zf$ represent the metallicity (with respect to solar) 
and the formation redshift of the stellar populations.
Uncoupled models ($\xi =-3w_Q$) are represented by the thick line.
}
\label{fig:dom}
\end{figure}

The results of our analysis are reported in figures~\ref{fig:quint} 
and \ref{fig:dom}. For each theoretical model we compare the ages 
obtained through equation~\ref{eq:age} with the observational data. 
Likelihood contours at the $68 \%$ (solid), 
and $95 \%$ (dashed) confidence levels are shown in 
the $\Omega_M-w_Q$ (uncoupled, Fig.~\ref{fig:quint})
and $\xi-w_Q$ (coupled, Fig.~\ref{fig:dom}) planes. 
Since the results depend also on $2$ extra
parameters involving the stellar populations, 
namely the metallicity ($Z_\star$) and the formation redshift
($z_F$) we have investigated different possible values 
for these two parameters.
In all the plots we have assumed a gaussian distributed 
prior on the Hubble parameter $h_0=0.72\pm 0.08$ as indicated 
by the Hubble Space Telescope Key project (Freedman et al. 2001).
In Fig.~\ref{fig:quint} we report the results of our analysis assuming
uncoupled dark energy models and for different choices
of $Z_\star$ and $z_F$. As we can see from the plots
a  degeneracy is present in the tradeoff between $w_Q$ and
$\Omega_M$ (higher values of $|w_Q|$ are consistent with a higher
$\Omega_M$) and no strong independent constraint can be 
obtained on these parameters. 
Furthermore, there is a strong and similar dependence
on formation redshift and metallicity. 
Namely, an increase in $z_F$ or $Z_\star$
shifts the likelihood contours towards higher values
of $\Omega_M$. A model with 
$\Omega_M=1$ and $w_Q=0$, for example, is excluded at high
significance when ``reasonable'' values of $z_F=2$ and
$Z_\star=Z_\odot$ are assumed, while an increase in 
formation redshift or metallicity pushes this model back
to the $68 \%$ confidence level. On the other hand, a low
value of $\Omega_M$ can be used as a way to revert the argument:
if we assume a ``concordance'' model with
$\Omega_M=0.3$ and $w_Q=-1$, then high values of the
formation redshift and the metallicity of the stellar
populations in bright cluster ellipticals are
disfavoured. 
Marginalizing over $z_F$ and $Z_\star$ gives the upper limits
$w_Q < -0.35$ and $\Omega_M<0.66$ at $68 \%$ C.L..
These limits even if less stringent are in good agreement with
 those obtained from observations of luminosity distances of
type~Ia supernovae.

In figure~\ref{fig:dom} we plot the contours in the $\xi-w_Q$ planes 
fixing $\Omega_M=0.3$. The thick diagonal line represents
uncoupled models.
The figure shows that even if  $\Omega_M$ is kept fixed,
no strong limit on $w_Q$ can be placed.
We cannot disentangle the degeneracy between $w_Q$ and $\xi$
if low metallicities ($Z_\star\sim Z_\odot$) or late formation
epochs ($z_F\sim 2$) are assumed.
However, coupled models ($\xi\neq -3w_Q$) are not favoured 
in our analysis, unless one is willing to have 
high $z_F$ and $Z_\star$. We should emphasize that $z_F$
is treated as a free parameter that determines the ages
of the stellar populations. Structure formation studies
for the models explored in this paper would provide
further constraints on the allowed region of parameter space.
Nevertheless, the models with $Z_\star=1.4Z_\odot$ are disfavoured in the light 
of the results from high redshift SNIa (Dalal et al. 2001) which 
rule out models with $w_Q>-0.5$ regardless of $\xi$ at the $99\%$ C.L.
Even models with $Z_\star =1.2Z_\odot$ would be hard to reconcile
with the SNIa data if the formation redshift is too high.

Figures~\ref{fig:quint} and \ref{fig:dom} illustrate the fact 
that an accurate
estimate of the metallicity along with a combined analysis
with high redshift SNIa would pose strong constraints on
dark energy parameters. High metallicities 
in massive ellipticals are expected if we assume
a scenario with a highly efficient star formation in which 
most of the ejecta from stars is recycled
back into subsequent generations of stars. Wind models
for elliptical galaxy formation support a mass-metallicity
relation which is caused by the potential well of the galaxy
which prevents significant gas outflows in massive galaxies 
thereby achieving high metallicities
(Larson 1974; but see Silk 2003). Massive elliptical
galaxies such as those targeted in this study should have
solar or even supersolar average metallicities. However,
the current data does not allow us to set a stringent 
constraint on $Z_\star$.  Hence, in addition 
to advancing our knowledge of galaxy formation and evolution,
accurate measurements of stellar metallicities in galaxies at
moderate and high redshift will have a very significant
impact on estimates of the cosmological parameters.

The likelihood regions shown in the
figure are in good agreement and/or consistent with those obtained by the
analysis of high redshift SNIa (Dalal et al. 2001) for most of the
models explored in this paper. 
Furthermore, our decoupled models (figure~\ref{fig:quint})
agree well with the estimates from globular cluster
ages (Krauss \& Chaboyer 2001).
Hence, the study of the colours of the stellar populations in 
moderate redshift elliptical galaxies deserves a place 
of its own as an estimator of cosmological parameters.

\section*{Acknowledgments}
It is a pleasure to thank Rachel Bean, Licia Verde 
and Joe Silk for useful comments. AM is supported by PPARC,
DT-V is funded by a Scatcherd Scholarship.

\end{document}